\documentclass[amsmath,showpacs,twocolumn,aps,prl]{revtex4}
\usepackage{bm}
\usepackage{graphics}
\begin{document}
\title{Intersubband Edge Singularity in Metallic Nanotubes}

\author{E.~G.~Mishchenko}
\affiliation{Department of Physics and Astronomy, University of Utah, Salt Lake
City, Utah 84112, USA}

\author{O.~A.~Starykh}
\affiliation{Department of Physics and Astronomy, University of Utah, Salt Lake
City, Utah 84112, USA}

\begin{abstract}
Tunneling density of states of both the massless and massive
(gapped) particles in metallic carbon nanotubes is known to have
anomalous energy dependence. This is the result of coupling to
multiple low-energy bosonic excitation (plasmons). For both kinds of
particles the ensuing effect is the suppression of the density of
states by electron-electron interactions. We demonstrate that the
optical absorption between gapless and gapped states is affected by
the many-body effects in the opposite way. The absorption
probability is enhanced compared with the non-interacting value and
develops a power-law frequency dependence, $A(\omega) \propto
(\omega-\Delta)^{-\gamma}$, where $\gamma \approx 0.2$ for typical
nanotubes.
\end{abstract}
\maketitle

{\it Introduction}. Energy spectrum of metallic carbon nanotubes (MNT) has massless
band  electrons propagating with velocity $v=8\times 10^5$ m/s. Despite
degeneracy of the spectrum at the Fermi level the backscattering between
the left- and right-movers is suppressed due to the orthogonality of
the two species. This makes MNT an ideal application for the
Luttinger Liquid (LL) theory \cite{LM,DL,H,G,review}. Observations of MNTs'
low-energy properties \cite{McEuen1,Dekker,Ishii} are consistent with the
picture of LL. According to
the latter the eigenmodes of the interacting system are bosons, while electrons are represented by
coherent combinations of infinite number of those modes. This leads to strong modification
of the electron spectral properties close to the Fermi level. A step
in the distribution function is replaced with the
interaction-dependent power law behavior. Furthermore, the tunneling
density of states becomes energy dependent, $\nu (\epsilon) \propto
\epsilon^\alpha$. The exponent $\alpha=(1-g)^2/2Ng$ depends on  the
total number of channels $N$, and the effective coupling constant
$g=v/u$ is determined by the velocity $u$ of a collective charged
mode -- plasmon.

\begin{figure}
\resizebox{.45\textwidth}{!}{\includegraphics{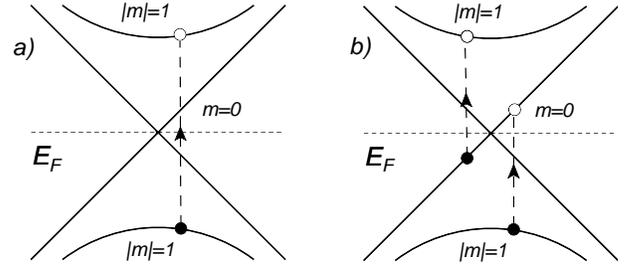}}
\caption{\label{fig1} Processes responsible for the optical absorption in metallic nanotubes. a) Polarization along the NT axis: only transitions with no change of the angular momentum, $m\to m$, are allowed (except $0\to 0$). The $E_{11}$ transition is indicated by the dashed line, the transition threshold $\omega=2\Delta$. b) Polarization perpendicular to the NT axis: $m\to m \pm 1$ is required. The processes involving right-moving massless particles are indicated. At zero temperature $T=0$ the $E_{01}$ and $E_{10}$ transitions are contributing to the optical absorption. The transition threshold $\omega=\Delta$.}
\end{figure}

In addition to the linear gapless states MNTs have subbands with the
nonzero angular momentum along the NT axis, see Fig.~\ref{fig1}. The
lowest subband has a gap $\Delta =v/R$ from the Fermi level that
depends on the tube radius $R$. In the absence of many-body effects
the corresponding density of states  has a one-dimensional van Hove
singularity. However, as demonstrated by Balents \cite{B}, the
interaction with massless electrons decreases the density of states
of the massive particle in a way reminiscent of LL supression, $\nu
(\epsilon) \sim (\epsilon -\Delta)^{-1/2+\beta}$, with
$\beta=(1-g^2)^2/2Ng$. The underlying physics of this
non-perturbative modification is indeed the same as in the case of
LL: coupling to multiple low-energy excitations. Surprisingly, while
technical advances of Ref.~\cite{B} have been used to further the
theory of one-dimensional systems beyond conventional LL approach
\cite{Glazman}, its physical implications for nanotubes have not
received due attention.

While the edge singularity uncovered in Ref.~\cite{B} could
potentially be probed by tunneling measurements or via x-ray
absorption, it remains unexplored if the underlying many-body
physics can be seen with the less experimentally intricate
techniques, in particular in the optical absorption. The main
difference comes from the fact that an optical transition happens
between {\it two states} strongly affected by the interaction with
LL, in contrast to transition from a deep core band, as is the case
in the x-ray absorption. Thus knowledge of a single-particle Green's
function is insufficient to address this question.

In the present paper we demonstrate that the solution to this problem depends strongly on the polarization of the incident electromagnetic field. In the
 case of {\it longitudinal} polarization the dipole optical transitions with the change of the azimuthal quantum number are forbidden.
 The allowed transitions, e.g.\ $E_{11}$ shown in Fig.~\ref{fig1}a), are {\it not affected} by the interactions with low frequency plasmons.
   The physical reason for this is the cancellation of the many-body effects in  the  propagators of the massive electron and hole states by the vertex corrections (which describe plasmon-mediated interactions in the final state of absorption transition).
Interaction of the electron and the hole suppresses their
probability to be at the same point via the (essentially
single-particle) physics of Sommerfeld factor, leading to strong
{\it suppression} from the free particle density of states
\cite{OT}. In particular,  for a short-range interaction, $A_{\parallel}(\omega) \propto
\sqrt{\omega-2\Delta}$ \cite{W}.

 To the contrary, transitions with
the change of the angular momentum, see Fig.~\ref{fig1}b), are
allowed for the {\it transverse} polarization of electric field. For
these transitions the cancellation does not occur and Coulomb
interaction leads to a singular deviation of the absorption spectrum
from the non-interacting model. According to the latter absorption
has a step-like threshold at $T=0$: e.g. for the $E_{01}$
transition, $A_\perp(\omega) \propto \int dp \delta
(\Delta+\frac{p^2}{2m}+v|p|-\omega) \propto \Theta(\omega-\Delta)$.
By taking electron-electron interactions into account we obtain the
divergent power law behavior,
\begin{equation}
\label{main_result}
A_\perp (\omega)\propto \frac{\Theta(\omega-\Delta)}{(\omega-\Delta)^\gamma}, ~~~~\gamma=\frac{2-g-g^3}{2N}.
\end{equation}
Since $g<1$ the effects of Coulomb interaction result in the {\it enhancement} of the optical absorption, unlike the x-ray absorption which is suppressed by the many-body effects. Below we derive our main result Eq.~(\ref{main_result}).

{\it Intersubband transitions.} Electronic band structure of carbon
nanotubes \cite{SDD} follows from the underlying two-dimensional
spectrum of graphene, $\varepsilon({\bf p})=\pm v |{\bf p}|$. The
components of the quasimomentum ${\bf p}$ are measured from the
corresponding Dirac points in the first Brillouin zone. In case of
rolled-up graphene sheets  the circumferential momentum $p_y$ is
quantized giving a set of one-dimensional subbands. For
``metallic'' folds some cuts pass through the Dirac points,
so that $p_y=m/R$, and the resulting spectrum consists of the
subbands,  $\varepsilon_m(p)=\pm v\sqrt{p^2+m^2/R^2}$, classified by
the angular momentum quantum number, $m=0,\pm 1, \pm 2, ...$; for
$(n,n)$ armchair tubes the NT radius $R=3na/2\pi$ where $a=1.4~
\text{\AA} $ is the distance between carbon atoms. The following
Hamiltonian describes the interaction of band electrons with the
external electric field of frequency $\omega$ polarized
perpendicularly to the MNT axis (we set $\hbar=1$),
\begin{equation}
\hat H_0=-iv\sigma_x\frac{\partial}{\partial x}-\frac{iv}{R}\sigma_y
\frac{\partial}{\partial \theta} +\frac{evE_0}{\omega}\hat \sigma_y \cos{\theta}\sin{\omega t},
\end{equation}
where $\theta=y/R$ is the circumferential angle. We now consider the
optical absorption corresponding to a transition from a massless
($m=0$) right-moving state to a massive particle ($m=1$) in the
first excited subband close to its bottom ($p\ll
1/R$), as indicated by the left dashed line in Fig.~\ref{fig1}b):
\begin{equation}
 \frac{\hat \psi_R(t,x)}{\sqrt{4\pi}} \left(
\begin{array}{c} 1
\\ 1\end{array} \right) \to \frac{\hat \Psi(t,x)}{\sqrt{4\pi}}
\left( \begin{array}{c} 1 \\ i\end{array} \right)e^{i\theta},
\end{equation}
where the operators $\hat \psi_{R,L}(t,x)$ and $\hat \Psi(t,x)$
describe the one-dimensional propagation of massless and massive
particles respectively. The probability of intersubband absorption
per unit length of MNT
\begin{equation}
\label{aperp}
 A_\perp(\omega)=\frac{\pi Ne^2 v^2 E_0^2}{4\omega^2} ~
{\cal V}(\omega)
\end{equation}
is determined by the following correlation function,
\begin{equation}
\label{correlator}
{\cal V}(\omega)=\int\limits_{-\infty}^\infty \frac{dt}{2\pi} dx e^{i\omega t} \langle [\psi^\dagger_R (t,x)\hat \Psi (t,x),\hat \Psi^\dagger (0,0)\hat \psi_R (0,0) ]\rangle
\end{equation}
Note that the overall coefficient in Eq.~(\ref{aperp}) also takes into
account  the transitions (all of which have the same
probability) to states with $m=-1$ as well as transitions from the
left-moving states, and also includes the total channel degeneracy $N$.
The correlator (\ref{correlator}) calculated for free
electrons yields the Golden Rule probability and reproduces
the step-like threshold discussed in the introduction.

{\it Electron-electron interaction}. Coulomb interaction in
quasi-one-dimensional wires has the form,
\begin{equation}
\hat H_{i}=\frac{1}{2} \int \frac{dq}{2\pi} V(q) \hat n(q) \hat n(-q),
\label{eq:Hint}
\end{equation}
where $\hat n$ is the operator of the total electron density.
Neglecting backscattering ensures that the formalism of the
Luttinger Liquid can be used. Backscattering and Umklapp scattering
amplitudes, $\sim e^2a/R$, are suppressed by virtue of the large
radius of a NT, $R\gg a$ \cite{KBF}. For ungated MNT $V(q)
=-2e^2\ln{|q|R}$, where $q\ll 1/R$. In case of a metallic gate
located a distance $d$ away, the geometry we assume here, the
logarithm is cut-off, $V(0)=2e^2\ln{d/R}$.

Coulomb interaction is screened by
the electron-hole excitations in the metallic ($m=0$) subbands, and is
given by the RPA dynamic propagator
\begin{equation}
\label{rpa}
V(0) \to U(\omega,q)=V(0)\frac{\omega^2-q^2v^2}{(\omega+i\eta)^2-q^2u^2},
\end{equation}
The poles of the expression (\ref{rpa}) correspond to collective eigenmodes of LL, plasmons, propagating with the velocity $u=v\sqrt{1+NvV(0)/\pi}\equiv v/g$. Plasmons are accompanied by electric field and the corresponding scalar potential can be expressed via the plasmon creation $\hat a_q^\dagger$ and annihilation  $\hat a_q$ operators,
\begin{equation}
\label{el_field} e\hat \phi(t,x)=u(1-g^2)\sum_q \sqrt{\frac{\pi
|q|}{2gN}}\Bigl[ \hat a_q e^{-i|q|ut+iqx} +c.c \Bigl],
\end{equation}
to be in agreement with Eq.~(\ref{rpa}).
In the bosonization scheme plasmons $\hat a_q$ represent total charge mode
of the system, while the remaining $N-1$ modes $\hat{b}_{i q}$ are
charge neutral (they account for spin and/or band degeneracy) and propagate with the Fermi velocity,
$\hat H =u\sum_q |q| \hat a_q^\dagger \hat a_q +\sum_{i=1}^{N-1}v\sum_q |q|
\hat b_{iq}^\dagger \hat b_{iq}$.

For the calculation of the correlator (\ref{correlator}) we need the
bosonized expression for the electron operators. Electrons in
gapless subbands are coherent combinations of bosonic modes, $\hat \psi_{R,L}(t,x)=\frac{1}{\sqrt{2\pi R}} \hat U_{R,L} e^{\hat
k_{R,L}(t,x)}$, where $\hat U_{R,L}$ are fermionic counting
operators;  the ultraviolet
momentum cut-off $1/R$ is being set by the NT radius. The phase operators are given by,
\begin{eqnarray}
\label{massless}
\hat k_{R,L} (t,x)=\sqrt{\pi}\sum_{q}\frac{1\pm g~ \text{sgn}~
q}{\sqrt{2gN|q|}}\Bigl[ \hat a_q e^{-i|q|ut+iqx} -c.c \Bigl]
\nonumber\\ + \sqrt{2\pi}\sum_{i=1}^{N-1}\sum_{q}\frac{\Theta(\pm
q)}{\sqrt{N|q|}}\Bigl[ \hat b_{iq} e^{-i|q|vt+iqx} -c.c \Bigl],
\end{eqnarray}
where the upper/lower sign is for the right/left-moving
electrons respectively.

 {\it Massive particle.} We utilize a simple eikonal approach to
 describe electrons belonging to the upper subband. We first demonstrate how the results of Ref.~\cite{B} are recovered with it.
 Schr\"odinger equation, describing
 interaction of the massive particle close to the bottom of the subband with the fluctuating electric field (\ref{el_field}) has the form,
\begin{equation}
\label{schroed} \left(i\frac{\partial}{\partial t}-\Delta
+\frac{1}{2\mu} \frac{\partial^2}{\partial x^2} \right)\hat
\Psi(t,x)=e\hat \phi(t,x) \hat \Psi(t,x),
\end{equation}
 where the effective mass is $\mu =1/vR$. For the heavy particle the solution can be obtained in the form,
\begin{equation}
\label{psi}
\hat \Psi(t,x)=e^{\hat K (t,x)}\hat \Psi^{(0)}(t,x).
\end{equation}
Here $\hat \Psi^{(0)}$ is the solution of Eq.~(\ref{schroed}) for
$\hat \phi =0$. Using the identity, $\partial_t e^{\hat K}=
(\partial_t K +\frac{1}{2}[K,
\partial_t K]) e^{\hat K }$, and {\it neglecting} spatial derivatives of the phase
$\hat K$, we obtain the following expression for the latter,
\begin{eqnarray}
\label{phase}
\hat {K} (t,x)&=&(1-g^2)\sum_q \sqrt{\frac{\pi }{2Ng|q|}}\Bigl[ \hat a_q e^{-i|q|ut+iqx} -c.c \Bigl] \nonumber\\
&&+it \frac{\pi u(1-g^2)^2}{2Ng} \sum_q 1.
\end{eqnarray}
The last term while formally divergent simply represents the
renormalization of the energy gap $\Delta$. This can be verified
directly by a perturbation calculation to the lowest order in the
dynamic interaction $U(\omega,q)$. The averaging over fermionic and
bosonic operators can now be performed independently in the
calculation of the massive particle's Green's function:
\begin{eqnarray}
\label{green's}
G_>(t,x)&=&G_>^{(0)}(t,x)\langle e^{\hat K
(t,x)} e^{-\hat K (0,0)}\rangle \nonumber\\
&&=-i R^\beta \sqrt{\frac{\mu}{2\pi t}} \frac{ e^{-i\Delta
t+i\mu x^2/2t}}{(u^2t^2-x^2)^{\beta/2}},~~~~~
\end{eqnarray}
where $\beta=(1-g^2)^2/2Ng$, in agreement with Ref.~\cite{B}. 

{\it Edge singularity.} Using the electron operators for massless (\ref{massless}) and
massive (\ref{psi})-(\ref{phase}) states we can now express the correlator (\ref{correlator}) through the bosonic average,
\begin{eqnarray}
\label{s}
{\cal V}(\omega)=&&\frac{1}{2\pi R} \int\limits_{-\infty}^\infty dt \int\limits_{-\infty}^\infty dx e^{i\omega t} G_>^{(0)}(t,x)
\nonumber\\  &&\times \langle e^{-\hat k_R(t,x)} e^{\hat K(t,x)} e^{-\hat K(0,0)} e^{\hat k_R(0,0)} \rangle.
\end{eqnarray}
As follows from Eq.~(\ref{green's}) the spatial integral in Eq.~(\ref{s}) converges on distances $x\sim \sqrt{t/\mu}$.
Since $t\sim 1/(\omega-\Delta)$ we observe that $x\ll vt,~ ut$. It is therefore sufficient to set $x=0$ in the correlation function of bosonic operators in Eq.~(\ref{s}). After straightforward calculation the latter is found to be $(R/v t)^{1-\gamma}$ with $\gamma$ defined in Eq.~(\ref{main_result}).
Here, as well as in the massive fermion's Green's function (\ref{green's}), it is assumed that $t \to t-i\eta$,
a complex plane is cut along the positive half of the imaginary $t$-axis and  the chosen branches assume real values on the negative half of the imaginary $t$-axis.

Performing now the $x$-integral in Eq.~(\ref{s}) and then the time
integral by deforming  the contour to follow the sides of  the
branch cut   we obtain for $\omega>\Delta$,
\begin{equation}
\label{v}
 {\cal V}(\omega)= \frac{1}{2\pi v \Gamma(1-\gamma)} \left( \frac{\Delta}{\omega-\Delta } \right)^\gamma.
\end{equation}
 For $\omega<\Delta$ the contour can be closed in the lower half-plane,
 and, because there are no singularities there, the function ${\cal V}(\omega)$ vanishes.
 When the interaction is absent, $\gamma \to 0$ and the function $ {\cal V}(\omega)$ approaches the one-dimensional density of states, ${\cal V}(\omega)=\Theta(\omega-\Delta)/2\pi v$ of the massless electrons.
For a MNT $N=4$. Assuming $R\approx 1$ nm, and thus, $\Delta =0.5$ eV, and $d\approx 50$
nm, the Coulomb coupling constant $g\approx 0.2$, which gives
$\gamma=0.2$.

It is worth noting that we also derived our results
\eqref{main_result} and \eqref{v} using  the formalism of
Ref.~\cite{B} as well as with the help of the unitary rotation
described in Ref.~\cite{KM}.

{\it Discussion.} The physical origin of the singularity (\ref{v})
is the non-perturbative interaction with multiple low-energy plasmon
excitations.  As external photon is being absorbed, excitation of an
electron and a hole is impeded by the creation of the virtual
plasmons. This is reflected in the suppression of each particle's
density of states. The interaction in the final state (vertex
corrections), to the contrary, {\it facilitates} propagation of the
electron-hole pair \cite{Fur}. In the LL description fluctuating
plasmon field amounts to the additional phase. When both particles
belong to massive subbands (e.g. in the $E_{11}$ transitions) the
phases accumulated by the electron and the hole are {\it opposite}
(due to opposite charges) and thus cancel each other. Formally this
can be seen from Eq.~(\ref{s}): in that case the phase $\hat
k_R(t,x)$ has to be replaced with $\hat K (t,x)$ and the bosonic
average drops out. Thus, the absorption lineshape is not affected by
many-body effects for the longitudinal polarization.

In case of $E_{01}$ absorption the electron and the hole belong to
{\it different} subbands. Since they propagate with unequal
velocities the phases accumulated from {\it the same} fluctuating
electric field are different \cite{Doppler}. The vertex corrections
do not exactly cancel the self-energy contributions and the
absorption line is modified. What is surprising about the spectrum
(\ref{v}) is that the enhancement by vertex corrections {\it
dominates}. This leads to the overall increase in the optical
response.

The increase of the transition probability by interactions is known
to occur in the x-ray edge problem in conventional metals
\cite{Mahan}. If the interaction is weak enough excitonic
contribution dominates, but with the increasing interaction strength
the subleading quadratic corrections begin to suppress absorbtion
via orthogonality catastrophe mechanism. We emphasize that in our
problem the enhancement occurs for any value of the interaction
strength $g$. Moreover, as opposed to the x-ray edge problem, in our
case both particles involved acquire non-trivial interaction-induced
dynamics.

One can make an interesting connection with the phase diagram of the
two-subband quantum wire studied in Ref.~\cite{KM}. It was found
there that for sufficiently small density of electrons in the second
(higher) subband {\em relative} charge fluctuations between the two
subbands are gapped. This correlated state is driven by
pair-tunneling processes which are strongly enhanced by the
inter-subband forward scattering (contained in \eqref{eq:Hint}). In
our case the second subband is empty but it remains true that
optically excited particles there are strongly correlated with
charge fluctuations in the lower (Luttinger liquid) subband. This
correlation, encoded in the phase factors of \eqref{s}, is the
reason for enhanced response \eqref{v}.

{\it Effects of finite doping and temperature} emphasize the
difference between $E_{01}$ and $E_{11}$ transitions. Because
typical  energy gaps $\Delta \sim 1$ eV the population of the gapped
subbands changes insignificantly at room temperatures or small
levels of doping the shape of $E_{11}$ should not change
appreciably. The behavior of the $E_{01}$ absorption is to the
contrary very sensitive due to  participation of gappless states.
The behavior at $E_F\ne 0$ is easy to infer from Fig.~\ref{fig1}b):
for small dopings the lineshape is simply given by
$\frac{1}{2}[A_\perp(\omega+E_F)+A_\perp(\omega-E_F)]$, and two
separate thresholds (with the relative  shift of $2E_F$) appear.

To consider finite $T$ effects it is sufficient to do this only for
the bosonic operators in  Eq.~(\ref{correlator}) since fermionic
Green's function $G_>$ is not affected at $T\ll \Delta$ and
$G_<\approx 0$. For $T\sim \omega -\Delta >0$ the  modification of
the lineshape is standard for LL problems: $A_\perp \propto
[\text{max}(T,\omega)]^{-\gamma}$. Additionally, subgap absorption
($\omega< \Delta$) appears for finite temperatures: the bosonic
average in Eq.~(\ref{s}), $(\pi TR/v\sinh{[\pi Tt]})^{1-\gamma}$,
now develops branching points in the lower half-plane
at $t=-i/T,~-2i/T,..$ The asymptotic behavior at $\Delta-\omega\gg
T$ is determined by the first of those points, ${\cal
V}(\omega)\propto \exp{(-\frac{\Delta-\omega}{T})}$. The interaction
strength
only affects the pre-exponential factor.

{\it Summary.} We have predicted singular power law enhancement of
the optical absorption between gapless and the first gapped
subbands of a single-wall metallic nanotube for the perpendicular polarization
of the incident radiation. The enhancement is the result of coupling
of both states to multiple plasmon excitations of MNT. For typical
MNTs the corresponding exponent is $\gamma \approx 0.2$ and depends only
weakly on the NT radius. Note that while the singularity should be present in multiwall NTs as well, the larger number of channels $N$ and disorder scattering from intrinsic incommensurability would be detrimental for its experimental observation.

{\it Acknowledgments.} Useful discussions with M. Raikh, L. Balents,
and A. Imambekov are gratefully acknowledged. This work was supported
by DOE, Office of Basic Energy Sciences, Grant No.~DE-FG02-06ER46313
(E.M.) and by the NSF Grant No. DMR-0808842 (O.S.).


\begin{thebibliography}{50}
\bibitem{Doppler} This is the manifestation of the Doppler effect: indeed, when formally, $u \gg v$  there is no
difference in the phase acquired by a ``fast'' massless particle
compared with the ``slow'' massive one as both the massless  and
masive particles are slow on the scale of plasmons. As a result both
exponents approach each other, $\alpha \approx \beta$.
\end{thebibliography}
\end{document}